\documentclass{sf2a-conf2024}
\usepackage{graphicx}
\usepackage{hyperref}
\usepackage[]{natbib}  
\usepackage{epstopdf}
\usepackage{amsmath}	
\usepackage{amssymb}
\usepackage{bm}
\usepackage{xcolor}

\def\BibTeX{{\rm B\kern-.05em{\sc i\kern-.025em b}\kern-.08em
    T\kern-.1667em\lower.7ex\hbox{E}\kern-.125emX}}
\bibpunct{(}{)}{;}{a}{}{,}  

\newcommand{\bn}{\bm\nabla}
\newcommand{\Pm}{\mathrm{Pm}}
\newcommand{\Ek}{\mathrm{Ek}}

\newcommand{\Le}{\mathrm{Le}}
\newcommand{\Lep}{\mathrm{Le_p}}

\newcommand{\edr}{E_\mathrm{dr}}
\newcommand{\Mp}{M_\mathrm{p}}

\newcommand{\Lec}{\mathrm{Le_c}}

\definecolor{db}{HTML}{191586}
\definecolor{vi}{HTML}{5E1111}

\begin{document}

\TitreGlobal{SF2A 2024}


\title{Interactions between tidal flows and magnetic fields in stellar/planetary convective envelopes}

\runningtitle{Magnetised tidal flows}

\author{A. Astoul}\address{School of Mathematics, University of Leeds, Leeds LS2 9JT, UK} 

\author{A. J. Barker$^1$} 

\setcounter{page}{237}


\maketitle

\begin{abstract}
Stars and gaseous planets are magnetised objects but the influence of magnetic fields on their tidal responses and dissipation rates has not been well explored. We present the first exploratory nonlinear magnetohydrodynamic (MHD) simulations of tidally-excited waves in incompressible convective envelopes harbouring an initial dipolar magnetic field. Simulations with weak magnetic fields exhibit tidally-generated differential rotation in the form of zonal flows (like in the purely hydrodynamic case) that can modify tidal dissipation rates from prior linear predictions. Moreover, tidal waves and zonal flows affect the amplitude and structure of the magnetic field, notably through creation of toroidal fields via the $\Omega$-effect. 
In contrast, simulations with strong magnetic fields feature severely inhibited zonal flows, due to large-scale magnetic stresses, excitation of torsional waves, or magnetic instabilities. We predict that the different regimes observed for weak and strong magnetic fields may be both relevant for low-mass stars when using turbulent values of the magnetic Prandtl number.
\end{abstract}



\section{Introduction}
Tidal interactions are the main driver of orbital and rotational evolution in compact stellar and exoplanetary two-body systems. Solar-like (low-mass) stars and giant gaseous planets feature convective envelopes in which waves, such as (magneto-)inertial waves restored by the Coriolis acceleration (and Lorentz force), can be tidally excited. Their dissipation contributes significantly to the angular momentum exchange in such systems. This is particularly true for fast rotators (young stars or Jupiter-like planets), since the (frequency-averaged) tidal dissipation scales approximately with the inverse square of the rotation period in linear theory \citep[e.g.][]{O2014}.

Very few global non-linear studies have been performed to explore the tidal response and its dissipation in convective envelopes of rotating stars and planets \citep[e.g.][]{T2007,FB2014,AB2022,AB2023}, and none with a magnetic field. Nevertheless, magnetic fields are ubiquitous in low-mass stars, as revealed by spectropolarimetry, which probes the large-scale magnetic fields at their surfaces, and as predicted by 3D MHD simulations of convective dynamos \citep[e.g.][]{BB2017}. Furthermore, although the frequency-averaged tidal dissipation when inertial waves are excited may not differ from linear hydrodynamical predictions \citep[e.g.][]{LO2018}, the nature of tidal waves and the mechanisms of their dissipation can be very different when considering a magnetic field \citep[see also][]{AM2019}. This is why we study here the interplay between tidal flows and magnetism in 3D non-linear simulations of rotating stellar convection zones, building upon our prior hydrodynamical studies in \cite{AB2022} and \cite{AB2023}, hereafter referred to as Paper~I and Paper~II, respectively. In these works, we found that non-linear self-interactions of tidally forced inertial waves induce cylindrical-like differential rotation (also called zonal flows). This differential rotation is particularly strong for thin shells, high tidal amplitudes, and low viscosities, where non-linear effects (including wave-wave, wave-zonal flow interactions and instabilities) play an important role. Indeed, in such cases, we found important deviations from linear predictions of tidal dissipation. In the following, we particularly focus on the effects of magnetism on the generation of differential rotation and how it modifies tidal dissipation rates, by varying the strength of an initial dipolar magnetic field, and the value of Ohmic diffusivity. 
%
\begin{figure}[ht!]
    \centering
    \includegraphics[trim=0.2cm 0.25cm 0cm 0.2cm,clip,width=0.46\linewidth]{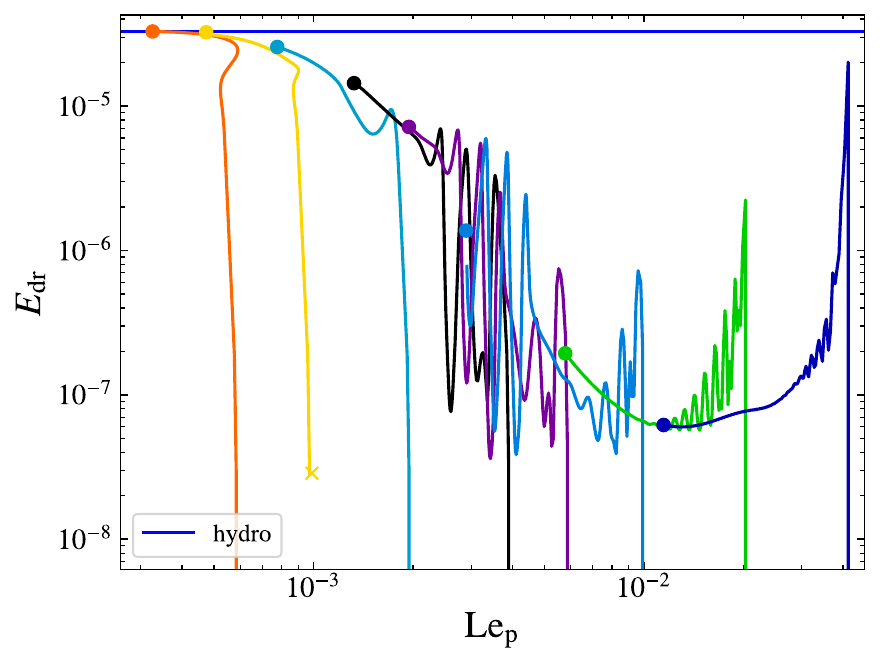}
    \includegraphics[trim=0cm 0.2cm 1cm 1.45cm,clip,width=0.48\linewidth]{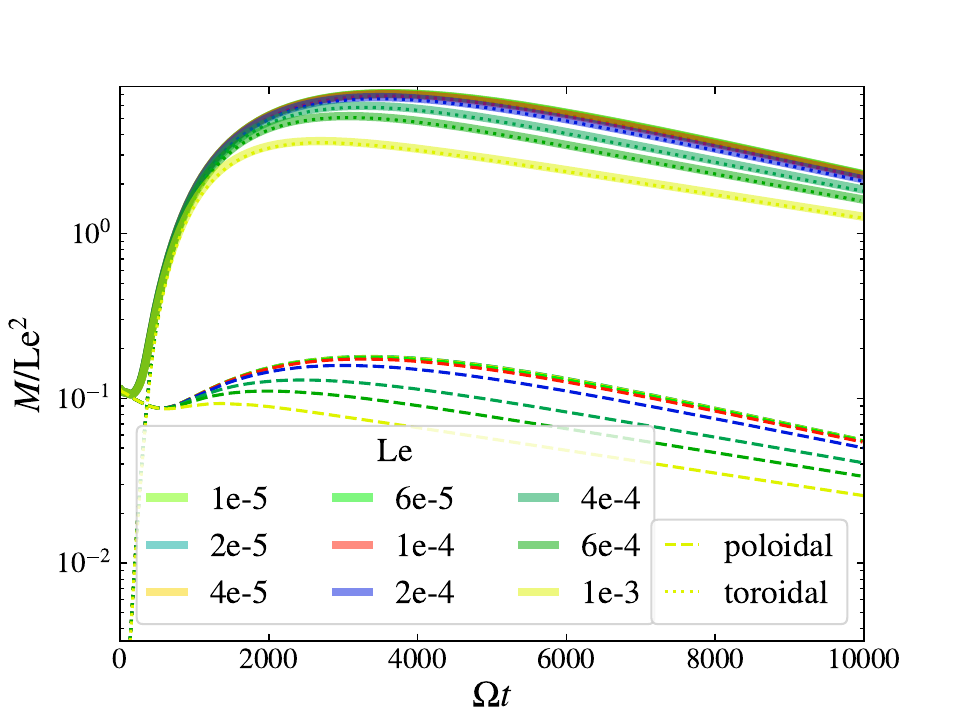}
        \vspace{-2ex}
    \caption{{\bf Left:} Energy in the differential rotation $\edr$ versus the evolving poloidal Lehnert number $\Lep$ for 8 simulations having different initial $\Le$ (in different colours). The value of $\edr$ at $t\approx10^4$ is indicated by a bullet and $\edr$ for $\Le=0$ is shown by a horizontal blue line.  
    {\bf Right:} Magnetic energy $M$ (solid lines) along with poloidal (dashed) and toroidal (dotted) components, all rescaled by $\Le^2$, versus time $\Omega t$, for simulations having different initial Lehnert numbers $\Le$.}
    \label{fig12}
        \vspace{-2ex}
\end{figure}
\section{MHD tidal model}
We build on the hydrodynamical and nonlinear tidal model described in Paper~I, to which we add an initial dipolar magnetic field of amplitude $B_0$. We solve the MHD equations for tidally-excited magneto-inertial waves in an incompressible and adiabatic convective shell. The size of the inner core (normalised to the total radius $R$) is fixed to $\alpha=0.5$, and the body rotates at a frequency $\Omega$ along $\bm e_z$, with a constant density $\rho$. 
The momentum, induction, and continuity equations for the magnetised tidal waves are then given:
\begin{subequations}
    \begin{align}
        \partial_t\bm u+2\bm e_z\wedge\bm u+(\bm u\cdot\bn)\bm u &=-\bn p+\Le^2(\bn\wedge\bm b)\wedge\bm b +\Ek\,\Delta\bm u+\bm f_\mathrm{t},\label{eq:mom}\\
        \partial_t\bm b &= \bn\wedge(\bm u\wedge\bm b)+\Ek\,\Pm^{-1}\,\Delta\bm b, \label{eq:ind}\\
        \bn\cdot\bm u&=0, \label{eq:solu}\\
        \bn\cdot\bm b&=0, \label{eq:solb}
    \end{align}
\label{eq:sys}%
\end{subequations} 
with $\bm u$, $\bm b$, $p$ the dimensionless velocity, magnetic field and pressure, using $R$, $\Omega^{-1}$, $B_0$ as units of length,  time, and magnetic field, respectively. For the tidal waves, we adopt stress-free and impenetrable\footnote{Note that this is certainly \textit{not} assumed for the total tidal flow including the non-wavelike component, which satisfies the correct (linear) free surface condition \citep[e.g.][]{AB2022}.} boundary conditions for the velocity, and current-free (i.e.~insulating, $\bm e_r\cdot(\bn\wedge\bm b)=0$) boundary conditions for the magnetic field at both the inner and outer shells, which is also continuously matched to a potential field. The effective tidal forcing $\bm f_\mathrm{t}$ is defined in a similar way as in Papers~I \& II. In Eq.~(\ref{eq:mom}), we have introduced the Lehnert number $\Le=B_0/(\sqrt{\mu\rho}R\Omega)$ which is a measure of the magnetic field strength, the Ekman number $\Ek=\nu/(R^2\Omega)$, where $\nu$ is the (effective turbulent) viscosity, set to $\Ek=10^{-5}$ here \citep[motivated by mixing-length theory, e.g.][]{OL2007}, and the magnetic Prandtl number $\Pm=\nu/\eta$ where $\eta$ is the Ohmic diffusivity.
As in \citet{FB2014} and Papers~I \& II, we define the energy within the differential rotation $\edr$ to measure the strength of the emerging zonal flows, and the tidal viscous dissipation $D_\nu$.
\section{Global simulations of magnetised tidal flows}
\begin{figure}[ht!]
    \centering
    \includegraphics[trim=0cm 0.2cm 0cm 0.2cm,clip,width=0.48\linewidth,clip]{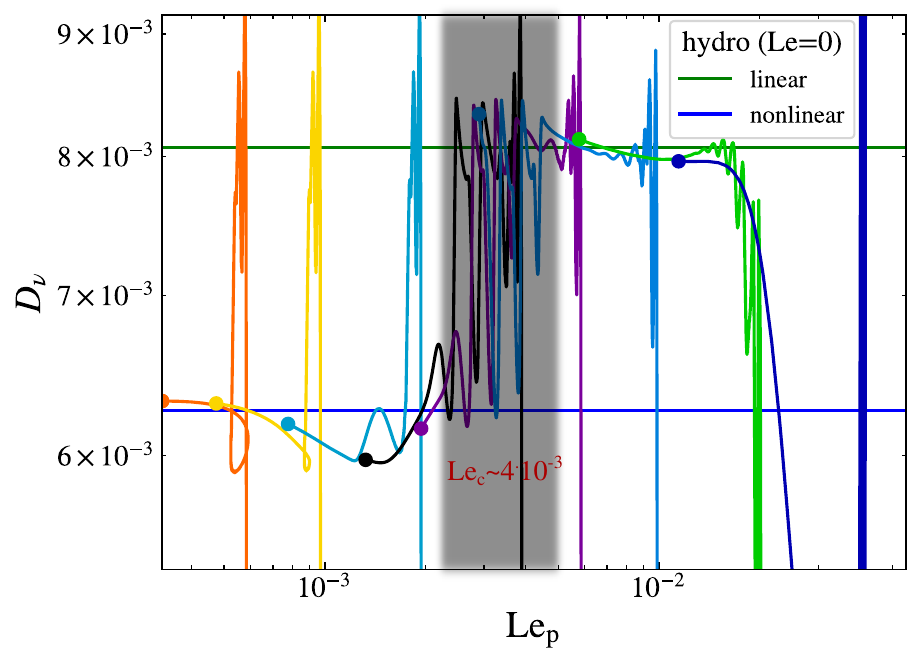}
    \includegraphics[trim=0cm 0.2cm 0cm 0.2cm,clip,width=0.48\linewidth,clip]{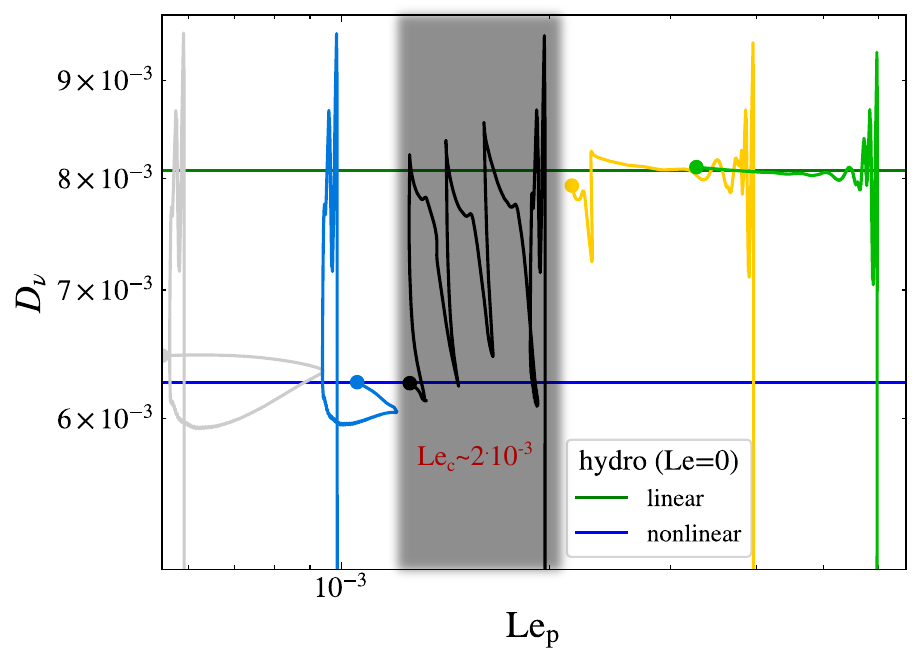}
    \vspace{-2ex}
    \caption{Tidal viscous dissipation $D_\nu$ versus evolving poloidal Lehnert number $\Lep$. Hydrodynamical ($\Le=0$) linear and nonlinear tidal dissipation rates are shown in horizontal green and blue lines. {\bf Left:} $\Pm=1$. {\bf Right:} $\Pm=2$.}
    \label{fig34}
    \vspace{-2ex}
\end{figure}
We use the pseudo-spectral code MagIC\footnote{\url{https://magic-sph.github.io/} has been modified to implement tidal interactions.} to solve the MHD system Eqs.~(\ref{eq:sys}) in a spherical shell.
At the beginning of the simulation ($t=0$), we impose an initial dipolar magnetic field \citep[defined exactly as in][]{LO2018}. Since the magnetic field is not self-sustained by convective motions (and usually decays), we also define $\Lep(t)=\left[3M_p/(\alpha^3\pi\left(1-\alpha^3\right))\right]^{1/2}$ such as $\Lep(t=0)=\Le$, to follow the evolution of the (volume-integrated) poloidal magnetic energy $\Mp$.
By running several simulations up to $t=10^{4}$ with various initial $\Le$ (in different colours), and $\Lep$ varying with time in the following figures, we observe two main regimes:
\begin{enumerate}
    \item For high Lehnert numbers $\Lep>10^{-3}$, the tidally-generated zonal flow is destroyed early in the simulations, as we can see in Fig.~\ref{fig12} (left panel). For example for $\Le=4\cdot10^{-4}$, $\edr$ first rises but is rapidly damped to reach a value more than two orders of magnitude lower than the corresponding hydrodynamic ($\mathrm{Le}=0$) value (blue horizontal line) in the steady state. This is due to Maxwell stresses arising from the strong initial dipolar magnetic field which quench differential rotation and thus counteract the effects of Reynolds stresses acting to generate zonal flows \citep[see e.g.][and references therein]{BB2017}. We also observe the generation of torsional Alfvén waves \citep[TOs; which are axisymmetric modes restored by the Lorentz force, e.g.][]{HN2023} resulting in oscillations of the zonal flow (seen primarily in $\edr$ but also in $D_\nu$ in Fig.~\ref{fig34}) with a frequency proportional to $\Lep$ \footnote{More details will be given in a forthcoming article.}. As $\Lep$ is decreased (and hence the poloidal magnetic energy), Maxwell stresses become weaker, so the differential rotation is stronger, and the Alfvén timescale becomes longer than the Ohmic diffusion timescale, which may explain why TOs are not observed for $\Le\lesssim10^{-3}$ as they are damped before propagating.
    \label{hiLe}
    \item For low Lehnert numbers $\Lep<10^{-3}$, the differential rotation is fully established in the simulations after a few thousand rotation times, and reaches the corresponding hydrodynamical value. Thus, zonal flows are strong enough to stretch the dipolar magnetic field lines to produce a toroidal axisymmetric component $\bm B_\Omega$, via the so-called $\Omega$-effect \citep{S1999}. Indeed, in Fig.~\ref{fig12} (right panel), we observe that the toroidal magnetic energy increases rapidly to exceed the poloidal magnetic energy. Soon after (for $t\gtrsim500$) the poloidal magnetic energy, which is initially dominated by the dipole, becomes predominantly quadrupolar. The stretching and twisting of the $m=0$ toroidal magnetic field by the $m=2$ wavelike tidal flow brings out a $m=2$ poloidal component of the magnetic field (through the term $(\bm B_\Omega\cdot\bn)\bm u_\mathrm{w}$ in the induction equation, where $\bm u_\mathrm{w}$ is the quadrupolar tidally-forced wavelike flow). This results in an increase in the poloidal magnetic energy, before both the poloidal and toroidal magnetic components peak and then decay due to the saturation of growth in differential rotation together with Ohmic diffusion. For $\Le\lesssim6\cdot10^{-5}$, a kinematic regime is reached where the Lorentz force can be neglected in the momentum equation, and the magnetic energies are proportional to $\Le^2$ with decreasing $\Le$.   
    \label{loLe}
\end{enumerate}
The two regimes described above are also visible in Fig.~\ref{fig34} displaying the viscous dissipation vs $\Lep$, with the grey zone emphasising the transition where powerful TOs are observed. When the zonal flows are washed out for high $\Lep$, it is interesting to see that the viscous dissipation is close to the rate predicted by the linear hydrodynamical simulation (horizontal green line), while it matches the nonlinear rate for lower $\Lep$ (horizontal blue line) when differential rotation becomes significant again. This is consistent with the fact that differential rotation and wave/zonal flow interactions in nonlinear simulations can strongly modify the tidal response and its dissipation, as found in \cite{BR2013} using a background cylindrical differential rotation and in Papers I \& II and \cite{FB2014} for a tidally generated one by nonlinear wavelike interactions.  
We stress that Ohmic dissipation in these simulations is quite small, mainly because the magnetic Prandtl number is $O(1)$ and not small \citep[as in][]{LO2018}, so the power injected by the tidal force is mainly balanced by viscous dissipation.

For higher magnetic Prandtl numbers (right panel in Fig.~\ref{fig34}), we also observe the same two magnetic regimes (as described above) around a critical Lehnert number $\Lec$. However, the transition between them is shifted towards lower Lehnert numbers, with $\Lec\approx4\cdot10^{-3}$ when $\Pm=1$, $\Lec\approx2\cdot10^{-3}$ when $\Pm=2$, and $\Lec\approx10^{-3}$ when $\Pm=5$ (not shown here). A possible explanation is that the Ohmic diffusion timescale becomes longer with increasing $\Pm$, so that the torsional Alfvén waves are less easily damped for similar magnetic field amplitudes at higher $\Pm$. The transition between the regimes dominated by, or with inhibited, zonal flows, also results from a subtle balance between the winding-up timescale (which depends on the shear of the differential rotation) and the Alfvén timescale \citep[similarly to][for instance]{JG2015}. In the convective envelopes of stars and planets, the microscopic magnetic Prandtl number is expected to be of the order of $10^{-2}$ or lower \citep[e.g.][]{BB2017}. For this tiny value of $\Pm$, we predict $\Lec$ to be much higher that what we quoted, although $\Ek$ should be considerably reduced if the microscopic viscosity is taken\footnote{$\Ek\sim10^{-12}$ for solar-like stars or $\sim10^{-18}$ for giant gaseous planets.}.
On the contrary, the turbulent magnetic Prandtl number is estimated to be of the order of one \citep[e.g.][]{KR2020}.  
Therefore, for Lehnert numbers in the range $[10^{-4},10^{-2}]$, as estimated for instance throughout the convective envelope of a one solar mass star \citep[][]{AM2019}, neither of the two regimes is excluded.
\section{Conclusions}
We have studied the non-linear interplay between tidal flows and magnetic fields using direct MHD simulations of rotating spherical convective shells. By varying the amplitude of an initial dipolar magnetic field (quantified by the Lehnert number $\Le$) in different simulations, and following the evolution of the poloidal component in each, we have highlighted two main regimes. While at high $\Le$ the tidally-driven zonal flow is destroyed, at low $\Le$ it survives and even affects the topology and the amplitude of the magnetic field. This is due to the stretching of first the dipolar magnetic component by the axisymmetric zonal flow to produce an azimuthal component, which is in turn stretched by the quadrupolar tidal flow to restore (temporarily) the poloidal component. The transition between the two regimes is associated with powerful torsional Alfvén waves, which cause significant oscillations in the zonal flows, which depend on the strength of the magnetic field, of the zonal flow and the value of Ohmic diffusivity. The tidal viscous dissipation rate also differs in the two regimes since it is strongly related to the presence and strength of the differential rotation. 

Although the zonal flow (when efficiently excited) affects the magnetic field components, Ohmic diffusion is still too strong for them to be self-sustained by tidal (zonal) flows in the simulations presented here. However, for higher magnetic Prandtl numbers we found evidence for magneto-rotational instabilities that periodically amplify the magnetic field. Describing these effects is beyond the scope of this proceeding and is the subject of an article in preparation.
\vspace{-2ex}
%
\begin{acknowledgements}
Funded by a Leverhulme Trust Early Career Fellowship to AA and by STFC grants ST/S000275/1 and ST/W000873/1. 
Simulations were undertaken on the DiRAC Data Intensive service at Leicester (STFC DiRAC HPC Facility \url{www.dirac.ac.uk}). 
\end{acknowledgements}
    \vspace{-2ex}
\bibliographystyle{aa}  
\bibliography{Astoul_S01} 
\end{document}